\documentclass[runningheads]{llncs}
\usepackage[T1]{fontenc}
\usepackage{graphicx, booktabs, url}
\usepackage{amsmath,amssymb,amsfonts}%
\usepackage{makecell}
\usepackage{multirow}
\usepackage{subcaption}
\usepackage[springer]{definition}
\definecolor{sblue}{HTML}{02BCD4}
\definecolor{sred}{HTML}{F44436}
\definecolor{spink}{HTML}{E91E62}
\definecolor{sgreen}{HTML}{8BC34A}
\definecolor{spurple}{HTML}{3F51B5}
\definecolor{slightgreen}{HTML}{CCDE3A}
\definecolor{sorange}{HTML}{FE9800}
\definecolor{sgolden}{HTML}{FFC108}
\newcommand{\ie}[0]{\textit{i.e.},~}
\newcommand{\eg}[0]{\textit{e.g.},~}

\newcommand{\modelname}[0]{{BrainNetMLP}}
\begin{document}
\title{BrainNetMLP: An Efficient and Effective Baseline for Functional Brain Network Classification}
\titlerunning{BrainNetMLP}
\author{Jiacheng Hou\inst{1}$^\dagger$ \and
Zhenjie Song\inst{1}$^\dagger$ \and
Ercan Engin Kuruoglu\inst{1}}
\authorrunning{Hou et al.}
\institute{Tsinghua Shenzhen International Graduate School, Tsinghua University, China
\\
\email{kuruoglu@sz.tsinghua.edu.cn}
}

% \author{Anonymized Authors}  %% Added for anonymized MICCAI 2025 submission
% \authorrunning{Anonymized Author et al.}
% \institute{Anonymized Affiliations \\
%     \email{email@anonymized.com}}
    
% %
\maketitle              % typeset the header of the contribution
\begin{abstract}
Recent studies have made great progress in functional brain network analysis by modeling the brain as a network of Regions of Interest (ROIs) and leveraging their connections to understand brain functionality and classify brain disorders. Various deep learning architectures, including Convolutional Neural Networks, Graph Neural Networks, and the recent Transformer, have been developed.  However, despite the increasing complexity of these models, the performance gain has not been as salient. Furthermore, the escalating model complexity will exacerbate the gap between theoretical research and real-world deployment.  
To mitigate this gap, we revisit the simplest deep learning architecture, the Multi-Layer Perceptron (MLP), and propose a pure MLP-based method, named \modelname\ for functional brain network classification. 
Specifically, \modelname\ incorporates a dual-branch structure to jointly capture spatial connectivity and temporal dynamics in spectral domain, enabling spatiotemporal feature fusion for precise classification. Besides, considering the fully-connected property of MLPs, we also propose an Edge-Degree Guided Pruning technique to remove redundant parameters of MLPs, further improving the efficiency.  We evaluate our proposed \modelname\ on two public and popular brain network classification datasets, the Human Connectome Project (HCP) and the Autism Brain Imaging Data Exchange (ABIDE). Experimental results demonstrate pure MLP-based methods can achieve state-of-the-art accuracy and efficiency, revealing the potential of MLP-based models as more efficient yet effective alternatives in functional brain network classification. The code is available at \url{https://github.com/JayceonHo/BrainNetMLP}.

\end{abstract}

\section{Introduction}
\let\thefootnote\relax\footnotetext{$\dagger$ denotes equal contribution.}
The human brain is an extremely complex system composed of numerous interacting regions. Investigating the organization and understanding the connectivity between different regions has long been a central goal of neuroscience, which is valuable in clinical analysis, diagnosis, and the corresponding treatment~\cite{fornito2016fundamentals,wig2011concepts,deco2011emerging}. For this reason, tremendous effort has been put into investigating the functional brain network. Among various brain networks, the Functional Connectivity (FC) network based on resting-state functional Magnetic Resonance Imaging (rs-fMRI) data is common and widely adopted. Given a specific atlas, Regions of Interest (ROIs) can be separated from rs-fMRI images and defined as nodes. The average Blood-Oxygen Level Dependent (BOLD) signal on each region is then extracted, and the statistical dependence between different regions is modeled as edges~\cite{simpson2013analyzing,smith2011network}. In this way, a brain functional network is built, and the analysis of it can effectively assist the prediction of neurological disorder~\cite{bannadabhavi2023community,kan2022brain}.
%and mental state~\cite{satterthwaite2015linked,weis2020sex}.

\begin{figure*}[tb]
    \centering
    \begin{subfigure}[t]{0.48\linewidth}
        \centering
        \includegraphics[width=\linewidth]{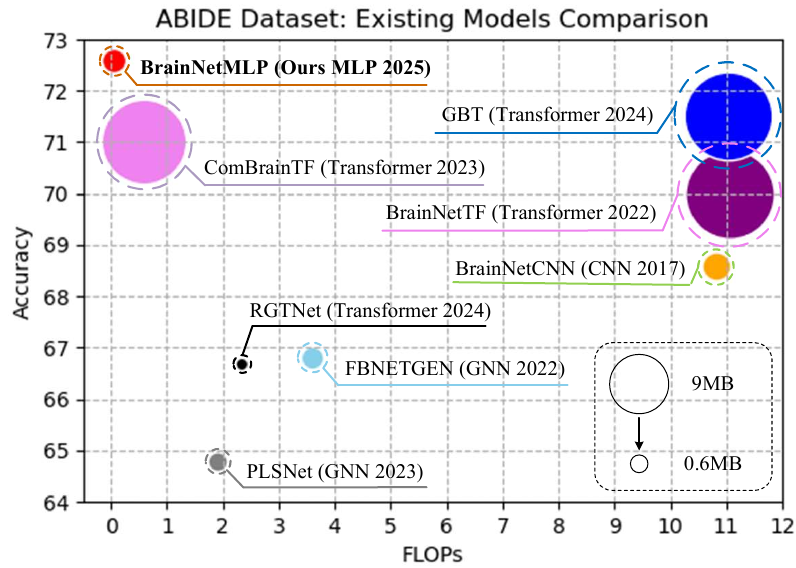}
        \label{fig:comp_abide}
    \end{subfigure}
    \hfill
    \begin{subfigure}[t]{0.48\linewidth}
        \centering
        \includegraphics[width=\linewidth]{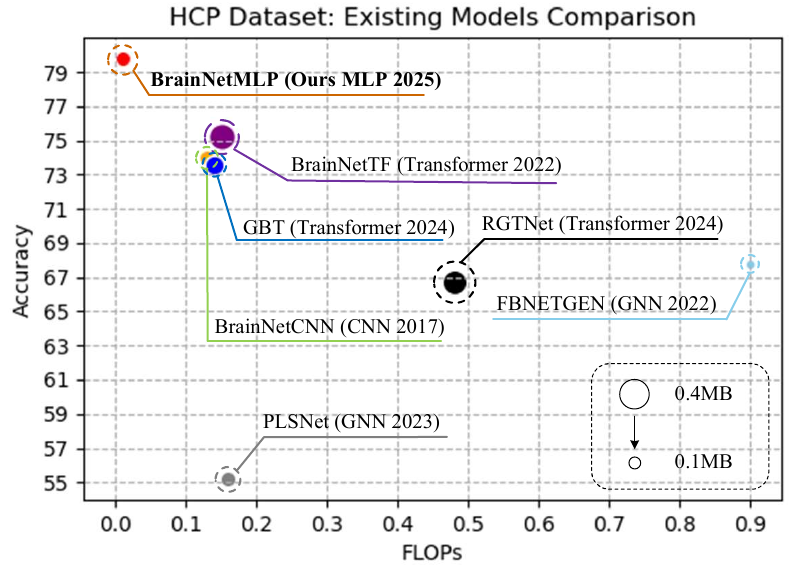}
        \label{fig:comp_hcp}
    \end{subfigure}
    \caption{Comparison of the existing state-of-the-art models and our proposed \modelname~model on ABIDE dataset and HCP dataset. The size of the circles corresponds to the size of the models.}
    \label{fig:complexity}
\end{figure*}

With the success of deep learning, various deep learning models have been adapted and applied to classify different brain functional networks, such as Convolutional Neural Networks (CNN), Graph Neural Networks (GNN), and Transformer~\cite{li2020braingnn, kawahara2017brainnetcnn, yan2024signal}. Among them, the Transformer has been favored mostly because of its strength in capturing long-range correlation of different ROIs~\cite{bannadabhavi2023community,kan2022brain,RGTNet}. However, as shown in Fig.~\ref{fig:complexity}, in comparison to the surge of model complexity, the performance gain of transformers is relatively marginal. Moreover, the increased complexity will hinder the real-world application.

Recently, the Multiple Layer Perceptron (MLP), a classical machine learning model, has been revisited, and a series of MLP-based models emerged as efficient yet effective alternatives to transformers in varying tasks~\cite{tolstikhin2021mlp,wang2022dynamixer,hu2021graph}.   
Compared to transformers, MLPs possess several inherent merits, such as lower computational complexity and fewer parameters, which are desired properties in brain network classification, especially when dealing with large-scale networks. Furthermore, owing to its fully connected structure, like transformers, the MLP model is also able to learn the underlying relationship between distant ROIs. 

Motivated by this, we propose a pure MLP-based method, named~\modelname. \modelname\ is a dual-branch architecture that incorporates two types of MLPs, named the Spatial Connectivity Mixer (SCMixer) and Spectrum ROI Mixer (SRMixer), to separately extract the discriminative features from FC and the filtered spectrum of the raw BOLD signals. The SCMixer leverages the symmetric property of the Pearson functional network to save computation and employs an MLP to globally learn ROI connectivity. Then, different from previous methods~\cite{gadgil2020spatio,bannadabhavi2023community,peng2024gbt}, we adopt spectral information in our designed SRMixer to utilize the temporal dynamics via spectral filtering and learning. The learned spatial and spectral features are fused to collaboratively classify the brain connectome. By exploiting the spatial connectivity of ROIs, their spectral dynamics, and the spatiotemporal correlation, high classification accuracy is attained. 

Additionally, considering the fully-connected property of MLPs, we propose an unstructural pruning technique, named Edge Degree Guided Pruning (EDGP) method. It prunes redundant parameters based on both the learnable weight and the edge degree calculated from the network to improve biological plausibility, yielding more comprehensive and effective pruning results.

\begin{itemize}
    \item We propose an efficient yet effective method, \modelname, for functional brain network classification, which is the first pure MLP-based method and demands much fewer computation (\ie 10$\times$ less FLOPs) and parameters (\ie 2$\times$ fewer parameters) compared with existing transformer models.
    \item \modelname\ not only leverages spatial connectivity information but also exploits effective spectral components of the brain network signal, thereby collaboratively capturing the spatiotemporal dynamics for precise analysis.
    \item To further improve efficiency, we propose an Edge Degree Guided Pruning method to produce more efficient variants of our \modelname.
\end{itemize}
% We evaluate our proposed \modelname\ method on two popular and public rs-fMRI datasets: the HCP and ABIDE datasets, which show state-of-the-art performance compared with existing larger models. This reveals the potential of the MLP structure for functional brain network classification. 

\section{Method}
\begin{figure*}
    \centering
    \includegraphics[width=0.85\linewidth]{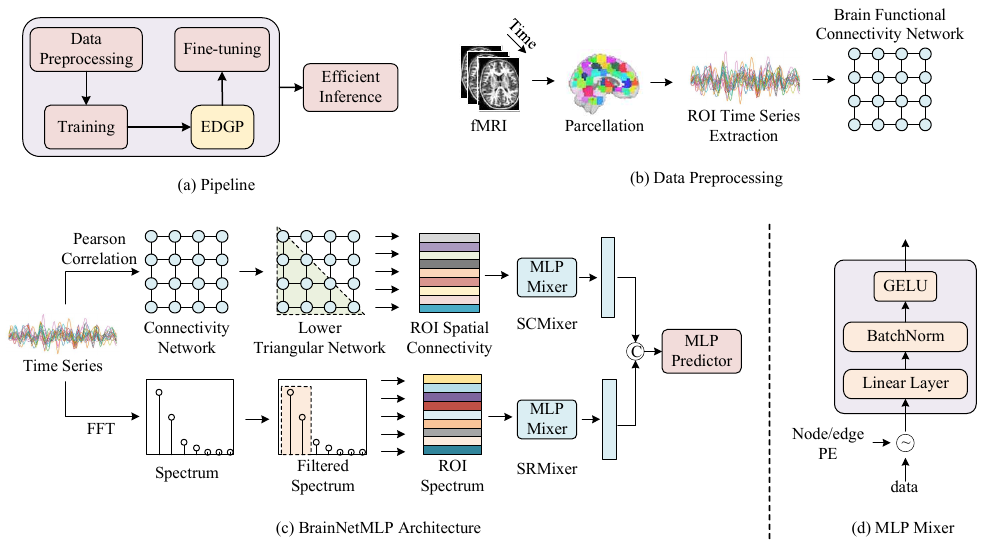}
    \caption{(a) the whole pipeline (b) the data preprocessing scheme (c) the structure of our \modelname\ (d) the structure of the MLP Mixer}
    \label{fig:pipeline}
\end{figure*}

% \subsection{Overview}
% \textbf{Problem definition.} 
In functional brain network classification, we begin by separating each rs-fMRI brain scan into $N$ ROIs. Subsequently, we construct the FC matrix $\mathbf{X}\in \mathbb{R}^{N\times N}$ based on the average BOLD time series $\mathbf{T}\in \mathbb{R}^{L\times N}$, where $L$ signifies the length of each time series. Each element within this matrix represents the Pearson correlation coefficient estimated between pairs of ROIs. The goal of the classification model is to predict the label $y$ (\eg the clinical diagnosis) from the given connectome $\mathbf{X}$ and time series $\mathbf{T}$ of the subject. The model pipeline is depicted in Fig.~\ref{fig:pipeline}, and we follow the pruning after training scheme~\cite{liu2018rethinking} to produce more efficient versions of \modelname\ for real-world deployment.

% \textbf{Model pipeline.} To solve this problem, we propose \modelname, the overall pipeline of which is depicted in Fig.~\ref{fig:pipeline}(a). Specifically, after data preprocessing, $\mathbf{X}$ and $\mathbf{T}$ are respectively fed to SCMixer and SRMixer for spatial and spectral feature extraction. Then, the extracted features are fused and sent to an MLP for prediction. After training, the model is pruned by our proposed TGP method to cut out redundant connections to obtain a more efficient model for inference.

\subsection{Spatial Connectivity Mixer}
Previous transformer models~\cite{RGTNet,bannadabhavi2023community,kan2022fbnetgen} directly treat the connectome $\mathbf{X}$ as a feature matrix. In these works, one dimension of $\mathbf{X}$ is regarded as tokens, and the other dimension is regarded as channels. Although the transformer structure allows token and channel mixing by the self-attention and feed-forward layer separately, we argue that this separate processing scheme is neither efficient nor necessary. In essence, each element in $\mathbf{X}$ represents the correlation between two ROIs, the meaning of which is quite different from the features of transformers in computer vision (\eg images and videos)~\cite{hou2024bidirectional,hou2024mcd, ji2025raformer} and natural language processing (\eg texts)~\cite{yu2019multimodal,zhang2021fast}. Thus, it is unnecessary to separately process elements of $\mathbf{X}$ by different modules. Based on this observation, the MLP structure becomes a good choice, as it can simultaneously and globally process all elements in the input. On the other hand, there is an important property of $\mathbf{X}$ that has long been ignored by previous transformer-based brain network classification models~\cite{bannadabhavi2023community,peng2024gbt,kan2022brain}, that is the symmetry of $\mathbf{X}$. The Pearson correlation is a commutable operation, which means $\mathbf{X}[i,j]=\mathbf{X}[j,i]$. Making use of this property can undoubtedly reduce the amount of computation. With these beliefs, in our designed SCMixer, we first extract the lower triangular (or upper triangular) matrix $\mathbf{X}_{lower}$ from $\mathbf{X}$. Then it is flattened to a one-dimensional vector $\boldsymbol{u}\in\mathbb{R}^{\frac{N(N+1)}{2}}$:
\begin{equation}
\begin{split}
    \mathbf{X}_{lower} &= \text{TriExact}(\mathbf{X}), \\
    \boldsymbol{u} &= \text{Flatten}(\mathbf{X}_{lower}),
\end{split}
\end{equation}
where ``TriExtract'' and ``Flatten'' represent the operations to extract the lower triangular matrix and flatten the matrix to a 1D vector.

In this way, the dimensionality of the to-be-processed feature (\ie $\frac{N(N+1)}{2}$) is reduced to nearly half of the whole connectivity matrix (\ie $N\times N$), which greatly saves computational resources. Besides, due to the symmetry property of $\mathbf{X}$, there is theoretically no loss of valid information. Then, a simple MLP Mixer (denoted by $f_{con}$) is applied to $\boldsymbol{u}$ for connectivity feature extraction:
\begin{equation}
    \boldsymbol{x}_c = f_{con}(\boldsymbol{u}  + \boldsymbol{p}_{edge}),
\end{equation}
where $\boldsymbol{p}_{edge}$ represents the positional embedding for each brain connectivity~\cite{yang2024brainmae}. Generally, $f_{con}$ plays the role of a learnable dimensionality reduction function, aiming to extract useful connectivity information and abandon non-discriminative connectivity information.

\subsection{Spectrum ROIs Mixer}
Besides the spatial connectivity information, an increasing number of works found that the network dynamics are also crucial~\cite{gadgil2020spatio,yan2021multi}, as the human brain network is evolving with the change of mental states. A straightforward way to capture the dynamics is to leverage the time series $\mathbf{T}$. However, due to the noise and variance in different collection sites~\cite{kan2022brain}, we find that directly using $\mathbf{T}$ for analysis would easily cause over-fitting. Hence, we choose to leverage the spectrum of $\mathbf{T}$, which is usually a more noise-robust feature representation. Specifically, in our developed SRMixer, we firstly apply the Fast Fourier Transform (FFT) on the temporal dimension of $\mathbf{T}$ as:
\begin{equation}
    \mathbf{S}_{real} = \text{FFT}(\mathbf{T}),
\end{equation}
where $\mathbf{S}_{real}$ represents half of the extracted spectrum (as the result of the FFT for real number is symmetric). Then, we implement the low-pass filtering on $\mathbf{S}_{real}$ to filter out the noise, as the noise is usually high-frequency components, which can be expressed by:
\begin{equation}
    \mathbf{S}_{real}^{low} = \mathcal{B}_{low}^k(\mathbf{S}_{real}),
\end{equation}
where $\mathcal{B}_{low}^k$ represents the low-pass filter, allowing only the first $k$ frequency components to pass. Then, the amplitude of $\mathbf{S}_{real}^{low}$ is sent to an MLP Mixer (denoted by $f_{roi}$) for spectrum ROIs learning as:
\begin{equation}
    \mathbf{X}_r=f_{roi}(|\mathbf{S}_{real}^{low}| +  \boldsymbol{p}_{rois}),
\end{equation}
where $\mathbf{X}_r\in\mathbb{R}^{\frac{L\times H}{2}}$ denotes the learned spectrum features; $\boldsymbol{p}_{rois}$ symbolizes the positional embedding for every ROI. Next, we average the spectrum features of different ROIs to obtain a more compact and comprehensive ROI feature representation as:
\begin{equation}
    \boldsymbol{x}_r = \text{Mean}(\mathbf{X}_r),
\end{equation}
where $\boldsymbol{x}_r$ represents the averaged spectrum features of ROIs.

\subsection{Feature Fusion and Pruning}
After obtaining $\boldsymbol{x}_c$ and $\boldsymbol{x}_r$ from the spatial connectivity and ROIs spectrum, we employ a simple non-linear projection layer (denoted by  $\mathcal{P}$) as a predictor to jointly estimate the label $\hat{y}$ as:
\begin{equation}
    \hat{y} = \mathcal{P}(\text{GELU}(\text{Norm}([\boldsymbol{x}_c,\boldsymbol{x}_r]))),
\end{equation}
where ``GELU'' and ``Norm'' represent the activation function~\cite{hendrycks2016gaussian} and normalization layers; $[~.~,.~]$ indicates the concatenation operation. Then, we employ the cross-entropy loss function as the objective function. We do not adopt additional loss functions to maintain the simplicity of \modelname\ in implementation and comparison. After training, we apply pruning to obtain more efficient variants.

Most of the parameters in \modelname\ stem from the SCMixer. Although we extract the triangular matrix to reduce the dimension of the feature to $\frac{N(N+1)}{2}$, when $N$ is large (\eg for large-scale brain networks), it will still lead to high complexity in computation and memory. To further decrease the complexity, we propose an unstructural pruning method, named Edge Degree Guided Pruning (EDGP), which computes an importance score $s_{i,j}$ for each parameter in the SCMixer based on the $\mathcal{L}_1$ norm and Edge Degree (ED) by:
\begin{equation}
    s_{i,j} = \| w_{i,j}\|_1 + ED(e_{i}) \cdot \lambda,
\end{equation}
where $\lambda$ is a coefficient to ensure the same numerical scale between $\| w_{i,j}\|_1$ and $ED(e_{i})$. $w_{i,j}$ and $e_i$ respectively denote the parameter in the linear layer of SCMixer and the element in $\boldsymbol{u}$. $ED(e_{i})$ is defined as the sum of node degrees connected by $e_{i}$. By this means, $s_{i,j}$ can be constrained by the brain network topology, rather than solely by the amplitude of the learned weight, as nodes with high degree are usually regarded as important~\cite{zhang2017degree, joyce2010new}. Based on the calculated importance weights, parameters $w_{i,j}$ with smaller $s_{i,j}$ can be cut to make room for storage and benefit speedup techniques (\eg sparse matrix multiplication).

\begin{table}[t]
	\caption{Quantitative performance comparison with baselines on HCP and ABIDE datasets (Mean$\pm$standard deviation). The best is shown in bold black.}
	\resizebox{\linewidth}{!}{
	\begin{tabular}{cccccccccccccc}
	\toprule
	\multirow{2}{*}{Method} & \multirow{2}{*}{Structure} & \multicolumn{4}{c}{Dataset: ABIDE} & \multicolumn{4}{c}{Dataset: HCP} \\
	\cmidrule(lr){3-6} \cmidrule(lr){7-10}
	& & Accuracy & AUCROC & Specificity & Sensitivity & Accuracy & AUCROC & Specificity & Sensitivity  \\
	\midrule
    VanillaCNN & CNN & 65.6$\pm$1.7 & 72.6$\pm$2.0 & 66.1$\pm$1.7 & 68.9$\pm$4.2 & 71.4$\pm$2.1 & 79.3$\pm$2.8 & 74.2$\pm$3.1 & 68.9$\pm$3.8\\
    BrainNetCNN~\cite{kawahara2017brainnetcnn} & CNN & 68.6$\pm$1.2 & 69.9$\pm$5.0 & 68.5$\pm$6.9 & 71.7$\pm$5.9 & 74.0$\pm$1.8 & 82.6$\pm$2.2 & 72.5$\pm$10.4 & \textbf{77.0$\pm$11.6}\\
    \midrule
     STGCN*~\cite{gadgil2020spatio} & GNN & - & - & -& -  & 77.7$\pm$3.2 & 86.9$\pm$2.5 & 82.5$\pm$4.3 & 72.2$\pm$3.3 \\
    BrainGNN~\cite{li2020braingnn} & GNN &  60.1$\pm$2.6 & 65.6$\pm$0.9 & 43.5$\pm$9.4 & \textbf{77.1$\pm$6.2} & 62.5$\pm$3.1 & 65.4$\pm$4.0 & 73.0$\pm$13.0 & 49.3$\pm$14.4 \\
    FBNETGEN~\cite{kan2022fbnetgen} & GNN &  66.8$\pm$1.9 & 74.8$\pm$1.0 & 72.5$\pm$5.1 & 62.2$\pm$7.0& 67.8$\pm$4.7 & 74.6$\pm$3.7 & 70.8$\pm$5.3 & 64.9$\pm$3.9 \\
    PLSNet~\cite{wang2023plsnet} & GNN & 64.8$\pm$1.1 & 71.8$\pm$2.3 & 67.1$\pm$2.5 & 64.2$\pm$2.7 & 55.2$\pm$4.6 & 58.5 $\pm$4.3 & 63.2$\pm$6.7 & 46.8$
    \pm$9.3\\
    \midrule
    BrainNetTF~\cite{kan2022brain} & Transformer & 70.0$\pm2.3$ & 79.1$\pm$1.1 & 68.3$\pm4.7$ & 72.7$\pm1.0$ & 75.2$\pm$2.1 & 82.5$\pm$1.6 & 81.0$\pm$6.1 & 69.6$\pm$6.4 \\
    ComBrainTF*~\cite{bannadabhavi2023community} & Transformer & 71.0$\pm$1.5 & 77.2$\pm$0.6 & 70.9$\pm$4.3 & 72.6$\pm$3.8 & - & - & -& - \\
    RGTNet~\cite{RGTNet} & Transformer & 66.7$\pm$2.8& 73.4$\pm$1.2 & 67.5 $\pm$1.8& 65.6$\pm$5.6 & 63.6$\pm$1.2 & 68.4$\pm$1.7 & 69.7$\pm$4.4 & 56.2$\pm$3.3 \\
    GBT~\cite{peng2024gbt} & Transformer &  71.5$\pm$2.1 & \textbf{79.2$\pm$0.4} & 69.5$\pm$4.3 & 72.9$\pm$3.5 & 73.6$\pm$2.3 & 81.4$\pm$2.6 & 80.1$\pm$7.2 & 66.7$\pm$9.0\\
	\midrule 
   \rowcolor{lightgray!20} Ours + 20\% pruning & MLP & 72.2$\pm$1.9 & 78.6$\pm$0.1 & 71.5$\pm$2.3 & 73.9$\pm$1.9 & \textbf{80.0$\pm$1.7} & 87.8$\pm$1.6 & \textbf{84.2$\pm$4.3} & 76.2$\pm$2.4 \\

   \rowcolor{lightgray!20} Ours + 50\% pruning & MLP & 72.4$\pm$1.7 & 79.1$\pm$0.1 & \textbf{72.9$\pm$2.7} & 72.5$\pm$2.6 & 76.8$\pm$2.9 & 86.0$\pm$2.7 & 84.1$\pm$6.2 & 68.9$\pm$13.2 \\
   
    \rowcolor{lightgray!20} Ours + no pruning & MLP & \textbf{72.6$\pm$1.7} & 78.4$\pm$0.1 & 72.5$\pm$2.7 & 73.7$\pm$1.4 & 79.8$\pm$1.7 & \textbf{88.1$\pm$1.5} & 83.1$\pm$3.4 & 76.8$\pm$4.1 \\
\bottomrule
	\end{tabular}}
    \tiny{* STGCN and ComBrainTF require fixed adjacency matrices and community priors, which are not provided in the ABIDE and HCP datasets.}
	\label{tab:performance}
\end{table}

\begin{figure*}[tp]
  \centering
  \begin{subfigure}{0.49\linewidth}
    \centering
    \includegraphics[width=\linewidth,clip, trim=0cm 0cm 0cm 0cm]{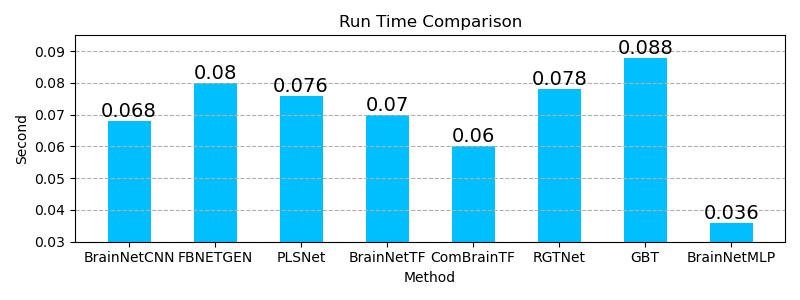}
    \caption{ABIDE dataset}
    \label{fig:abide_time}
  \end{subfigure}
  \begin{subfigure}{0.49\linewidth}
    \centering
    \includegraphics[width=\linewidth,clip, trim=0cm 0cm 0cm 0cm]{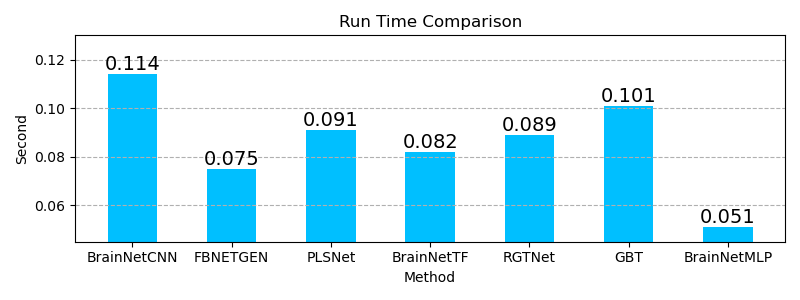}
    \caption{HCP dataset}
    \label{fig:hcp_time}
  \end{subfigure}
  \caption{Runtime comparison between our method and the other compared methods, for classifying a subject's rs-fMRI scans.}
  \label{fig:comp_time}
\end{figure*}

\begin{figure*}[tp]
  \centering
  \begin{subfigure}{0.49\linewidth}
    \centering
    \includegraphics[width=\linewidth,clip, trim=0cm 0cm 0cm 0cm]{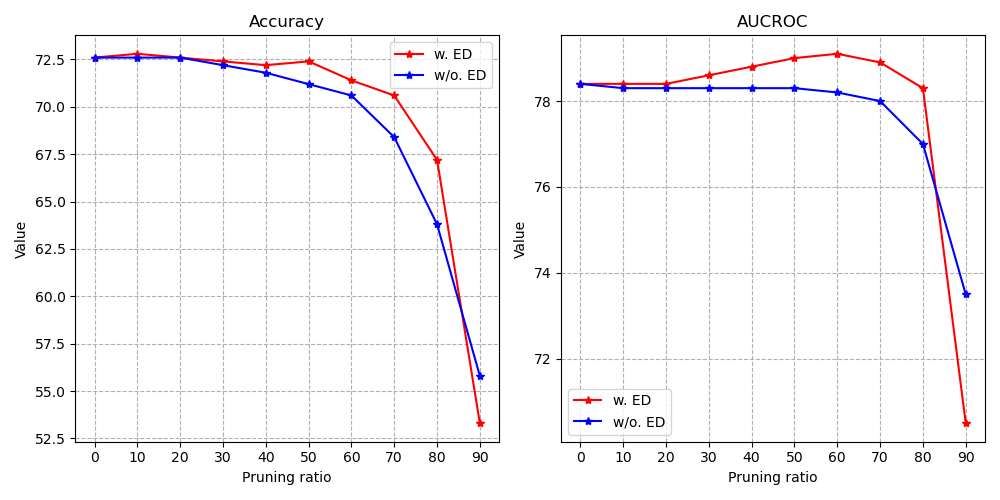}
    \caption{ABIDE dataset}
    \label{fig:abide_k}
  \end{subfigure}
  \begin{subfigure}{0.49\linewidth}
    \centering
    \includegraphics[width=\linewidth,clip, trim=0cm 0cm 0cm 0cm]{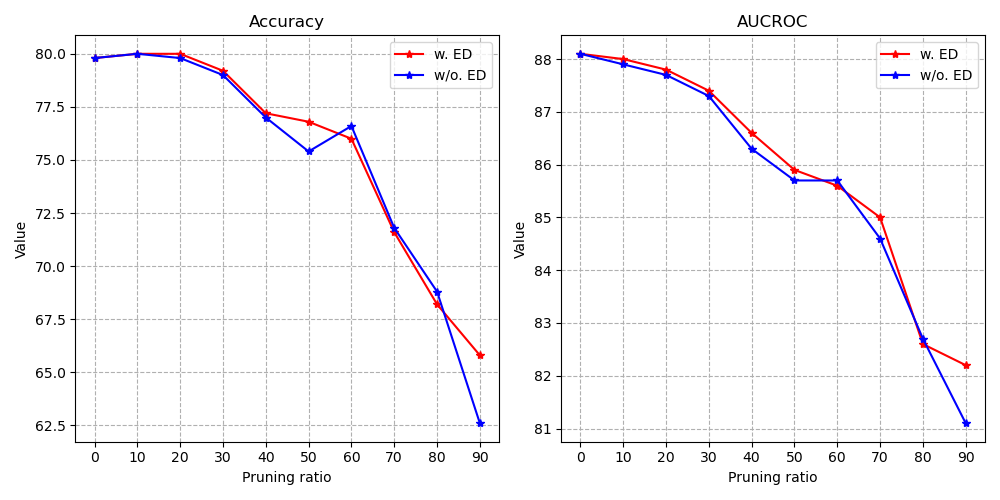}
    \caption{HCP dataset}
    \label{fig:hcp_k}
  \end{subfigure}
  \caption{Pruning results in different ratios with and without ED.}
  \label{fig:pruning}
\end{figure*}

\begin{table}[]
    \caption{Ablation study on the architecture of our proposed \modelname. }
    \resizebox{\textwidth}{!}{
    \begin{tabular}{lcccccccc}
    \toprule
    \multirow{2}{*}{Method}   & \multicolumn{4}{c}{Dataset: ABIDE} & \multicolumn{4}{c}{Dataset: HCP} \\
     \cmidrule(lr){2-5} \cmidrule(lr){6-9}
         &  Accuracy & AUCROC & Specificity & Sensitivity &  Accuracy & AUCROC & Specificity & Sensitivity\\
    \midrule
         w/o. SCMixer & 52.2$\pm$1.5 & 51.9$\pm$2.3 & 34.0$\pm$8.0 & \textbf{74.0$\pm$7.5} & 61.6$\pm$2.9 & 66.0$\pm$3.0 & 62.2$\pm$2.8 & 62.1$\pm$6.5\\
         w/o. SRMixer &  62.0$\pm$6.1 & 75.0$\pm$1.8 & \textbf{77.0$\pm$17.5} & 48.5$\pm$27.6 & 77.6$\pm$2.9 & 86.4$\pm$1.4 & 82.0$\pm$1.3 & 73.6$\pm$6.5\\
         w/o. filtering* & 71.8$\pm$1.5 & 78.4$\pm$0.3 & 71.7$\pm$1.6& 73.5$\pm$2.2 & 78.6$\pm$1.9 & 87.6$\pm$1.6 & 81.1$\pm$2.4 & 76.8$\pm$5.0\\
    \midrule
         \rowcolor{lightgray!20} Full model  & \textbf{72.6$\pm$1.7} & \textbf{78.4$\pm$0.1} & 72.5$\pm$2.7 & 73.7$\pm$1.4  & \textbf{79.8$\pm$1.7} & \textbf{88.1$\pm$1.5} & \textbf{83.1$\pm$3.4} & \textbf{76.8$\pm$4.1}\\
    \bottomrule
    \end{tabular}}
    \tiny{* filtering means only removing the low-pass filter $\mathcal{B}_{low}^k$ in SRMixer. The best is shown in bold black.}
    \label{tab:ablation}
\end{table}

\section{Experimental Results} 

\subsection{Quantitative comparison}
\textbf{Performance comparison.} We compare the performance quantitatively in Table~\ref{tab:performance}, where it can be seen that \modelname\ achieves state-of-the-art performance on both datasets. Specifically, on the ABIDE dataset, \modelname\ improves the accuracy by around 1.1\% and realizes comparative AUCROC compared with the second-best model. On the HCP dataset, \modelname\ yields a 2.1\% improvement in accuracy and a 1.2\% increase in AUCROC, observably outperforming the second-best model. This can be attributed to the additionally incorporated spectral information in \modelname, which has been ignored by the existing methods (\eg GBT and BrainNetTF). The experimental settings can be found in our supplementary material.\\
\textbf{Complexity comparison.} As compared in Fig.~\ref{fig:complexity}, the consumed computational cost (FLOPs) and the number of learnable parameters are all significantly fewer than the state-of-the-art transformers. Moreover, with the expansion of the network (\ie from 22 ROIs per network in the HCP dataset to 200 ROIs per network in the ABIDE dataset), unlike the rapid increase of complexity in the GBT (0.14G/0.21M $\rightarrow$ 11.04G/8.89M) and BrainNetTF (0.15G/0.4M $\rightarrow$ 11.06G/9.08M), our \modelname\ exhibits more advanced scalability (0.01G/0.14M $\rightarrow$ 0.06G/0.65M). The detailed FLOPs and the number of parameters can be found in our supplementary material.\\
\textbf{Runtime comparison.} Due to the model simplicity, as shown in Fig.~\ref{fig:comp_time}, the inference speed of our method is much faster than all other methods, and only one third/a half time consumed for classifying an functional brain network. Note that the functional brain network is estimated from rs-fMRI images spanning across a period of time (\eg 100 time steps in ABIDE and 1000 time steps in HCP), so the time spent for HCP dataset is longer than that of ABIDE dataset.

\subsection{Ablation Study}
\textbf{Ablation study on ED.} We test the model performance in accuracy and AUCROC under different pruning ratios (from 0\% to 90\%) with and without our added edge degree criterion. The results are shown in Fig.~\ref{fig:pruning}. It can be found that the pruning performance can be elevated with the aid of ED, which even produces the best performance at a pruning ratio of 50\% on the ABIDE dataset. This demonstrates the existence of spurious connections in the ABIDE dataset, and ED can assist in discovering these connections, while pure $\mathcal{L}_1$ norm cannot. Since only 22 important ROIs and their connections are used in the HCP dataset, the performance gain of ED is not obvious.\\
% \textbf{Interpretibility of \modelname.} Besides quantitative results, we also visualize how does \modelname\ make decisions on the ABIDE dataset. Particularly, we generate the saliency maps~\cite{simonyan2013deep} in Fig.~\ref{fig:brain} respectively based on the gradients of the SCMixer and SRMixer modules, which visualizes the salient connectivity/ROIs for ASD and HC subjects. From Fig.~\ref{fig:brain}, we can find the the major difference exhibited in the DMN and SMN, which is in accordance with other research on ASD~\cite{harikumar2021review}.
\textbf{Ablation study on model architecture.} To investigate the effectiveness of each module, we remove them from the full model. As reported in Table~\ref{tab:ablation}, on both datasets, no matter which component is removed, the overall performance is worse than the full model. Notably, on the ABIDE dataset, solely employing SCMixer or SRMixer can achieve the best specificity and sensitivity, which shows SCMixer and SRMixer can respectively help decreasing the rate of false alarms and miss detections. As a result, it can be observed that adaptively combining them complementarily improves the overall performance of the full model. More ablation studies can be found in the supplementary material.
\\
% \textbf{Ablation study on spectrum filter.} In SRMixer, we adopt a low-pass filter to remove invalid frequency components. From Table~\ref{tab:ablation}, without filtering, the accuracy on both datasets is reduced by around 0.8\% and 1.2\% separately. The hyper-parameter $k$ of the filter determines which frequency components are used for learning. The visualized spectrum and performance of different selections of $k$ are shown in Fig.~\ref{fig:spectrum}, where we can find selecting a suitable $k$ can keep valid frequency information and filter out noise, effectively boosting the performance.

\section{Conclusion}
In this paper, we propose the first MLP-only method, named \modelname, for functional brain network classification. Owing to the simplicity of the MLP structure, it is superior in efficiency and scalability, with more than one order of magnitude reduction in computational complexity compared with the state-of-the-art transformer models. Besides, we design a dual-branch architecture, which utilizes spatial connectivity and spectral features with adaptive feature fusion. As a result, \modelname\ produces advanced classification performance on the ABIDE and HCP datasets. Moreover, we propose a brain topology-aware pruning method to further save parameters for the real-world deployment. The experimental results illustrate the significant untapped potential of the MLP structure in functional brain network classification and outline an efficient yet effective alternative to the transformers.\\

\begin{credits}
\subsubsection{\ackname} This work is supported by Shenzhen Science and Technology Innovation Commission under Grant JCYJ20220530143002005, Shenzhen Ubiquitous Data Enabling Key Lab under Grant ZDSYS20220527171406015, and Tsinghua Shenzhen International Graduate School Start-up fund under Grant QD2022024C.

\subsubsection{\discintname} The authors have no competing interests to declare that are relevant to the content of this article.
\end{credits}

% \begin{credits}
% \subsubsection{\ackname} A bold run-in heading in small font size at the end of the paper is
% used for general acknowledgments, for example: This study was funded
% by X (grant number Y).

% \subsubsection{\discintname}
% It is now necessary to declare any competing interests or to specifically
% state that the authors have no competing interests. Please place the
% statement with a bold run-in heading in small font size beneath the
% (optional) acknowledgments\footnote{If EquinOCS, our proceedings submission
% system, is used, then the disclaimer can be provided directly in the system.},
% for example: The authors have no competing interests to declare that are
% relevant to the content of this article. Or: Author A has received research
% grants from Company W. Author B has received a speaker honorarium from
% Company X and owns stock in Company Y. Author C is a member of committee Z.
% \end{credits}

% ---- Bibliography ----
\bibliographystyle{splncs04}
\bibliography{ref}

\end{document}